\documentclass[pra, 12pt, showpacs, showkeys, eqsecnum, nofootinbib]{revtex4}
\usepackage{graphics}

\newcommand {\psig}{\Sigma_}
\newcommand {\pxi}{\Xi_}
\newcommand {\apsig}[1]{\langle \Sigma_{#1} \rangle}
\newcommand {\apxi}[1]{\langle \Xi_{#1} \rangle}
\newcommand {\apsigxi}[2]{\langle \Sigma_{#1} \Xi_{#2} \rangle}

\newcommand {\Tr}{{\mbox{Tr}}}

\begin{document}
\title{Maps and inverse maps in open quantum dynamics}
\author{Thomas F. Jordan}
\email[email: ]{tjordan@d.umn.edu}
\affiliation{Physics Department, University of Minnesota, Duluth, Minnesota
55812}

\begin{abstract}

Two kinds of maps that describe evolution of states of a subsystem coming from dynamics described by a unitary operator for a larger system, maps defined for fixed mean values and maps defined for fixed correlations, are found to be quite different for the same unitary dynamics in the same situation in the larger system. An affine form is used for both kinds of maps to find necessary and sufficient conditions for inverse maps. All the different maps with the same homogeneous part in their affine forms have inverses if and only if the homogeneous part does. Some of these maps are completely positive; others are not, but the homogeneous part is always completely positive. The conditions for an inverse are the same for maps that are not completely positive as for maps that are.  
For maps defined for fixed mean values, the homogeneous part depends only on the unitary operator for the dynamics of the larger system, not on any state or mean values or correlations. Necessary and sufficient conditions for an inverse are stated several different ways: in terms of the maps of matrices, basis matrices, density matrices, or mean values. The inverse maps are generally not tied to the dynamics the way the maps forward are. A trace-preserving completely positive map that is unital can not have an inverse that is obtained from any dynamics described by any unitary operator for any states of a larger system.

\end{abstract}


\maketitle

\section{Introduction}\label{one}

How extensively can open quantum dynamics be described by changes of density matrices made by linear maps? How much do properties of the maps reflect physical properties of the dynamics? Does the dynamics determine the form of the map? Does the dynamics determine whether the map has an inverse? Does an inverse map correspond to the reversed dynamics? Does it correspond to any dynamics at all? When we go from a map to its inverse are we still in the physical realm of operations that can be done to the system through dynamical interactions with other systems? Or are we only in the mathematical realm of maps? Can these questions be handled by using only completely positive maps? 

The most basic questions are about the connection between maps and dynamics. They are brought out here by the fact that the same open dynamics can be described by two different kinds of maps. These questions are put in sharper focus when inverse maps are considered. 

Quantum information theory has given emphasis to completely positive maps, and quantum error correction works with partial inverses that also are completely positive maps \cite{Schumacher96,Knill97,Nielsen97,Nielsen98,nielsen00}. Completely positive maps describe evolution of states of a system that interacts with another system but is initially not correlated with the other system. These maps can be engineered simply by bringing the uncorrelated systems together and letting them interact. More generally, evolution of states of a system in open quantum dynamics, a system that interacts with another system and initially may be correlated with the other system, can be described by maps that are not completely positive \cite{pechukas94,alicki95,pechukas95,stelmachovic01a,jordan04,jordan05b}. In particular, in quantum information processing, unwanted correlations will bring in maps that are not completely positive. Moreover, the inverse of a map, indeed the inverse of a completely positive map, generally is not completely positive.

Maps that are not completely positive can be used to learn things whose truth does not depend on the maps. For example, they have been used to describe how Lorentz transformations of spin depend on momentum for a particle with spin 1/2 and positive mass, find that every Lorentz transformation completely removes the information from a number of spin density matrices \cite{me81}, find that finite Lorentz transformations can produce maximal entanglement of both the spins and the momenta from separable states of two particles with finite momenta \cite{me82}, and find that the entanglement of two qubits can be increased from zero to maximal by an interaction on just one of them \cite{me84}. When the maps used as tools are set aside, the results about Lorentz transformations or open dynamics remain.

Things can happen in open quantum dynamics that are not described by completely positive maps. On the other hand, we will see here that working with maps that are not completely positive can bring in maps and properties of maps that are not connected with dynamics. They may not describe anything that can happen physically. No results will remain when the maps are set aside.  

Here we consider maps, completely positive or not, that describe evolution of states of a subsystem coming from dynamics described by a unitary operator for a larger system. Mathematically, this includes all trace-preserving completely positive maps; they all can be associated with unitary operators for larger systems \cite{nielsen00}. We consider two more general kinds of maps: those defined using only fixed mean values as map parameters \cite{jordan04,jordan05b}; and those defined using correlations \cite{stelmachovic01a}. We carefully describe and compare them. Each gives a correct description of the change in time of every state in its domain. The domain can be different for the two kinds of maps. We will see from examples that the forms of the two kinds of maps can be quite different in the same situation for the same unitary dynamics in the larger system. The conditions for a map to have an inverse are very similar for the two kinds of maps.

Maps that are not completely positive have limited domains. Does this make the conditions for an inverse more complicated for these maps than for completely positive maps? We look at each map in an affine form \cite{jordan04b} where the homogeneous part is a completely positive linear map. All the different maps with the same homogeneous part have inverses if and only if the homogeneous part does. The conditions for an inverse are the same for maps that are not completely positive as for maps that are. For maps for fixed mean values, whether there is an inverse depends only on the unitary operator for the dynamics of the larger system; it does not depend on any state, mean values, correlations, or absence of correlations. For maps for fixed correlations, whether there is an inverse depends  on the unitary operator for the dynamics of the larger system and on the state of the other part of the larger system; it does not depend on any correlations or absence of correlations between the two parts.

We will review, extend, and apply a framework \cite{jordan04,jordan05b} for describing open quantum dynamics as a map of density matrices that extends to a linear map of matrices, as a map of basis matrices, or as a map of mean values. Describing the map these different ways gives alternative statements of necessary and sufficient conditions for an inverse: that the homogeneous part of the affine map does not map any nonzero matrix to zero; that linearly independent basis matrices are mapped to linearly independent matrices; or that the homogeneous part maps the linear space of all matrices for the open system one to one onto itself, not into a smaller subspace. These conditions apply to both kinds of maps. For maps for fixed mean values, a necessary and sufficient condition for an inverse is that a linear map in a space of mean-value vectors that corresponds to the homogeneous part does not map any nonzero mean-value vector to zero, or that a multiple of the identity matrix is the only density matrix that the homogeneous part maps to a multiple of the identity matrix. For a qubit, this is the condition that the center is the only point in the Bloch sphere that is mapped to the center. For maps for fixed correlations, the condition is that the zero mean-value vector and another vector are not both mapped to the same vector, or that a multiple of the identity matrix and another density matrix are not both mapped to the same density matrix. For a qubit, this is the condition that the center and another point in the Bloch sphere are not both mapped to the same point.

For maps for fixed mean values, a necessary and sufficient condition for an inverse can be stated simply in terms of the unitary operator for the dynamics in the larger system: that it does not map all the matrices for the subsystem into a subspace of the linear space of matrices for the larger system that is linearly independent of a subspace of matrices for the subsystem. For maps for fixed correlations, an example will show that a condition for an inverse can not generally be stated in terms of the dynamics alone.

We consider an example of the open dynamics of one qubit that interacts with another. For both kinds of maps, we find that the homogeneous part is a completely positive map with an inverse that is not completely positive. The inverse map is not tied to the dynamics the way the map forward is. We look at the reason for this most closely for the map for fixed mean values. In that case, the evolution of states of the one qubit that comes from the dynamics of the two qubits going forward in time is described by a single map specified by a fixed set of mean values of quantities that involve the other qubit. The reversed dynamics of the two qubits gives many separate maps that take the states of the one qubit back in time. These maps generally are different for different states of the one qubit because they depend on mean values involving the other qubit that are changed by the dynamics so that at the later time they are different for different states of the one qubit. These separate maps do not join to form a single inverse map. The inverse map is not obtained this way. It does not come from the reversed dynamics the way the forward map comes from the forward dynamics. There is a similar disconnect of the reversed dynamics and the inverse map for fixed correlations. 

These are examples of a more general result. We show that a trace-preserving completely positive map that is unital can not have an inverse that is obtained from any dynamics described by any unitary operator for any states of a larger system.

Our results provide answers to the basic questions asked in the first paragraph. Open quantum dynamics generally can be described by changes of density matrices made by linear maps. We describe two different kinds of maps that can be used. Properties of the two kinds of maps can be quite different for the same dynamics. The form of the map as a whole is different. For one kind of map, the dynamics determines whether the map has an inverse; for the other kind it does not. An inverse map generally does not correspond to the reversed dynamics. It might not correspond to any dynamics at all. It does not, for example, for any trace-preserving completely positive map that is unital. Inverses generally exist only in the mathematical realm of maps; the connection to the physical realm is broken when inverses are taken.

\section{Framework}\label{two}

We use a framework that has been nearly completely developed already \cite{jordan04,jordan04b,jordan05b}. Consider two interacting quantum systems $S$ and $R$, both described by finite  matrices: $N$$\times$$N$ matrices for $S$ and $M$$\times$$M$ for $R$.
We use the basis matrices $F_{\mu 0}$, $F_{0 \nu}$ and $F_{\mu \nu}$ described previously \cite{jordan04,jordan05b}. The $F_{\mu 0}$ for $\mu = 0,1, \ldots N^2-1$ are $N^2$ Hermitian matrices for $S$ such that $F_{00}$ is $\openone_S$, the unit matrix for $S$, and
\begin{equation}
  \label{eq:f1}
  \Tr_S \left[ F_{\mu 0} F_{\nu 0} \right] = N \delta_{\mu \nu}.
\end{equation}
This implies that the $F_{\mu 0}$ are linearly independent, so every matrix for $S$ is a linear combination of the $F_{\mu 0}$. For example, the $F_{\mu 0}$ for $\mu = 1,2, \ldots N^2-1$ could be obtained by normalizing standard generators \cite{tilma02a} of $SU(N)$. The $F_{0 \nu}$ for $\nu = 0, 1, \ldots M^2-1$ are $M^2$ Hermitian matrices for $R$ such that $F_{00}$ is $\openone_R$, the unit matrix for $R$, and
\begin{equation}
  \label{eq:f2}
  \Tr_R \left[ F_{0 \mu} F_{0 \nu} \right] = M \delta_{\mu \nu}.
\end{equation}
Every matrix for $R$ is a linear combination of the $F_{0 \nu}$. We use notation that identifies $F_{\mu 0}$ with $F_{\mu 0} \otimes \openone_R$ and $F_{0 \nu}$ with $\openone_S \otimes F_{0 \nu}$ and let
\begin{equation}
  \label{eq:f3}
  F_{\mu \nu} = F_{\mu 0} \otimes F_{0 \nu}.
\end{equation}
Every matrix for the system of $S$ and $R$ combined is a linear combination of the $F_{\mu \nu}$.

We follow common physics practice and write a product of operators for separate systems, for example a product of Pauli matrices $\Sigma$ and $\Xi$ for two different qubits, simply as $\Sigma \Xi$, not $\Sigma \otimes \Xi$. Occasionally we insert a $\otimes$ for emphasis or clarity. 

The matrices $F_{\mu 0}$ for positive $\mu $ and $F_{0\nu }$ for positive $\nu $ are generalizations of Pauli matrices (and like the Pauli matrices they have zero trace). We use them to describe density matrices the way we use Pauli matrices to describe density matrices for qubits. If $\Pi$ is a density matrix for the system of $S$ and $R$ combined, then
\begin{equation}
  \label{eq:f4}
  \Pi = \frac{1}{NM}[\openone + \sum_{\alpha =1}^{N^2-1} \langle F_{\alpha 0} \rangle F_{\alpha 0} + \sum_{\mu =0}^{N^2-1}\sum_{\nu  =1}^{M^2-1} \langle F_{\mu \nu } \rangle F_{\mu \nu }]
\end{equation}
and the density matrix $\rho$ for $S$ is
\begin{equation}
  \label{eq:f5}
  \rho = \Tr_R \Pi = \frac{1}{N}[\openone_S + \sum_{\alpha=1}^{N^2-1} \langle F_{\alpha 0} \rangle F_{\alpha 0}]
\end{equation}
so that
\begin{equation}
  \label{eq:f6}
  \langle F_{\alpha \beta} \rangle = \Tr\left[ F_{\alpha \beta} \Pi \right]
\end{equation}
and in particular
\begin{equation}
  \label{eq:f7}
  \langle F_{\alpha 0} \rangle = \Tr_S \left[ F_{\alpha 0} \, \Tr_R \Pi \right] = \Tr_S \left[ F_{\alpha 0} \rho \right].
\end{equation}

If $U$ is a unitary matrix, then
\begin{equation}
  \label{eq:f8}
  U^{\dagger}F_{\alpha \beta } U = \sum_{\mu  =0}^{N^2-1} \sum_{\nu  =0}^{M^2-1}t_{\alpha  \beta  \,;\, \mu \nu }F_{\mu \nu }
\end{equation}
with the $t_{\alpha \beta  \,;\, \mu \nu }$ elements of a real orthogonal matrix \cite{jordan04}, so that $t^{-1}_{\alpha \beta  \,;\, \mu \nu }$ is $t_{\mu \nu  \,;\, \alpha \beta }$. Since $U^{\dagger} \openone U$ and $U \openone U^{\dagger}$ are $\openone$,
\begin{equation}
  \label{eq:f9}
  t_{00 \,;\, \alpha \beta} = \delta_{0\alpha} \, \delta_{0 \beta} \;, \quad t_{\alpha \beta \,;\, 00} = \delta_{\alpha 0} \, \delta_{\beta 0}.
\end{equation}

\section{Dynamics}\label{three}

We consider evolution described by a unitary matrix $U$ for the system of $S$ and $R$ combined. In the Heisenberg picture, each matrix $A$ that represents a physical quantity for $S$ is changed to $U^{\dagger}AU$. Its mean value is changed to $\langle U^{\dagger}A U \rangle $. These changes of mean values determine the change of the state of $S$. In particular, the state of $S$ is described by the mean values $\langle F_{\alpha 0} \rangle$ for positive $\alpha $, which determine the density matrix $\rho$ in Eq.(\ref{eq:f5}). Taking mean values in Eq.(\ref{eq:f8}) gives
\begin{equation}
  \label{Umv}
  \langle U^{\dagger}F_{\alpha 0} U\rangle  = \sum_{\mu  =1}^{N^2-1}t_{\alpha 0 \,;\, \mu 0}\langle F_{\mu 0}\rangle  + \sum_{\mu =0}^{N^2-1} \sum_{\nu =1}^{M^2-1}t_{\alpha 0 \,;\, \mu  \nu }\langle F_{\mu \nu }\rangle
\end{equation}
for $\alpha  = 1,2, \ldots N^2-1$. This is for one state of $S$ that is part of a state of $S$ and $R$ combined.

\section{Maps for fixed mean values}\label{four}

We would like to describe the open dynamics of $S$ with a single map that applies to a set of different states of $S$. We will see two ways this can be done. In Section VII, we consider maps for fixed correlations \cite{stelmachovic01a}. Here we describe maps for fixed mean values \cite{jordan04,jordan05b}.

We define a map $\hat {\Omega }$ of mean values $\langle F_{\alpha 0} \rangle$ for positive $\alpha $ by letting
\begin{equation}
  \label{hatF}
  \hat {\Omega }(\langle F_{\alpha 0} \rangle ) = \langle U^{\dagger}F_{\alpha 0} U\rangle 
\end{equation}
with the $\langle F_{\mu \nu } \rangle $ for positive $\nu $ in Eq.(\ref{Umv}) held fixed. We consider the $\langle F_{\mu \nu } \rangle $ for positive $\nu $ that participate in Eq.(\ref{Umv}) to be map parameters. They describe the effect of the dynamics of the larger system of $S$ and $R$ combined that drives the evolution of $S$. There may be $\langle F_{\mu \nu } \rangle $ for positive $\nu $ that are not map parameters; they have no effect on the change of states of $S$ because in Eq.(\ref{Umv}) they are with $t_{\alpha 0 \,;\, \mu  \nu }$ that are zero. Which $\langle F_{\mu \nu } \rangle $ are map parameters depends on $U$.

The map $\hat {\Omega }$ applies to all the states of $S$ described by mean values $\langle F_{\alpha 0} \rangle$ that are compatible with the fixed map parameters $\langle F_{\mu \nu } \rangle $ in describing a possible initial state of $S$ and $R$ combined. This set of states of $S$ is called the compatibility domain of the map \cite{jordan04,jordan05b}. For each state of $S$ in the compatibility domain, the state of $S$ and $R$ is described by a density matrix $\Pi_\Omega $ given by Eq.(\ref{eq:f4}) for the $\langle F_{\alpha 0} \rangle$ that describe the state of $S$ and the fixed map parameters $\langle F_{\mu \nu } \rangle $. 

The mean value $\langle A \rangle = \Tr_S[A\rho ]$ for a matrix $A$ for $S$ is changed to
\begin{equation}
  \label{eq:a2}
 \langle U^{\dagger}A U \rangle = \Tr[U^{\dagger}A U\Pi_\Omega ] = \Tr_S[ A \; \Tr_R[U \Pi_\Omega U^{\dagger}]].
\end{equation}
so the Schr\"{o}dinger picture for $S$ is that the density matrix $\rho$ for $S$ is changed to\footnote[1]{The notation is that for each map $X$ of matrices, the corresponding map of mean values is $\hat {X}$.}
\begin{equation}
  \label{eq:a3}
  \Omega (\rho ) = \Tr_R[ U \Pi_\Omega U^{\dagger}] = L(\rho) + K
\end{equation}
where
\begin{equation}
  \label{Ldef}
  L(Q) = \Tr_R[ U \, Q \, \frac{\openone_R}{M} U^{\dagger}] \nonumber 
\end{equation}
\begin{equation}
\label{eq:a4}
  K = \Tr_R[ U(\Pi_\Omega  - \rho \, \frac{\openone_R}{M} )U^{\dagger}] = \frac{1}{NM}  \sum_{\mu =0}^{N^2-1}\sum_{\nu  =1}^{M^2-1} \langle F_{\mu \nu } \rangle \Tr_R[UF_{\mu \nu }U^{\dagger}] .
\end{equation}

The $L$ part is a completely positive linear map that applies to any matrix $Q$ for $S$, density matrix or not. It has the property that $L(\openone_S)$ is $\openone_S$. The map $L$ depends on $U$ but does not depend on the state of $R$ or on mean values of any quantities that involve $R$. The matrix $K$ is the only part of $\Omega (\rho )$ that depends on the state of $R$ or on mean values that involve $R$. It does not depend on the state of $S$. It depends on $U$ and on the map parameters $\langle F_{\mu \nu } \rangle $ for positive $\nu $, but not on the $\langle F_{\alpha 0} \rangle$. With the same $K$, the Eq.(\ref{eq:a3}) defines a map $\Omega $ that applies to different density matrices $\rho$ representing different states of $S$.

Different map parameters $\langle F_{\mu \nu } \rangle$ specify different maps. Each map $\Omega $ applies to different states of $S$ described by different $\langle F_{\alpha 0} \rangle $. For each map $\Omega $ there is one $N$$\times$$N$ matrix $K$.

The map $\Omega $ of density matrices $\rho $, described by Eqs.(\ref{eq:a3})-(\ref{eq:a4}), extends to a linear map of all matrices for $S$. The map of the basis matrices $F_{\alpha 0}$ to
\begin{equation}
  \label{eq:l2}
  \Omega (\openone_S ) = \openone_S + NK, \; \; \; \; \quad \Omega(F_{\alpha 0}) = L(F_{\alpha 0})
\end{equation}
for positive $\alpha $ is equivalent to Eqs.(\ref{eq:a3})-(\ref{eq:a4}) for a set of $\rho $ described by Eq.(\ref{eq:f5}) with variable $\langle F_{\alpha 0} \rangle$. Since $K$ is the same for all $\rho$, it cannot come from the terms with variable coefficients $\langle F_{\alpha 0} \rangle$; it can only be part of $\Omega (\openone_S )$. For each matrix $Q$ for $S$, density matrix or not, the map is that
\begin{equation}
  \label{anyQ}
  \Omega (Q) = L(Q) + K \Tr_S (Q).
\end{equation}

The change of states of $S$ can be described equally well by the map $\hat {\Omega }$ of the mean values $\langle F_{\alpha 0} \rangle$ for positive $\alpha $. For each state, these $\langle F_{\alpha 0} \rangle$ are the components of a vector in a space of $N^2 -1$ dimensions. Consider the change of states of $S$ described by $L$. Because $L(\openone_S )$ is $\openone_S $, the map $\hat {L}$ for this change of states extends to a linear map of the space of mean-value vectors. To see this, we work out the equations that are needed. 

If mean values $\langle F_{\alpha 0} \rangle_1 $ are for a density matrix $\rho_1 $ and $\langle F_{\alpha 0} \rangle_2 $ are for a density matrix $\rho_2 $ in Eq.(\ref{eq:f5}), then the mean values for the density matrix
\begin{equation}
\label{eq:ld2}
\rho = p\rho_{1} + (1-p)\rho_{2},
\end{equation}
for $p$ between $0$ and $1$ are
\begin{equation}
\label{Fsum}
\langle F_{\alpha 0} \rangle = p\langle F_{\alpha 0} \rangle_1 + (1-p)\langle F_{\alpha 0} \rangle_2
\end{equation}
and, because
\begin{equation}
\label{Lrho}
L(\rho ) = pL(\rho_{1} ) + (1-p)L(\rho_{2} ),
\end{equation}
the mean values for $L(\rho )$ are\footnotemark[1]
\begin{equation}
\label{LFsum}
\hat {L}(\langle F_{\alpha 0} \rangle ) = p\hat {L}(\langle F_{\alpha 0} \rangle_1 ) + (1-p)\hat {L}(\langle F_{\alpha 0} \rangle_2 ).
\end{equation}
In particular, when $\langle F_{\alpha 0} \rangle_2 $ is zero for all nonzero $\alpha $, so
\begin{equation}
\label{Fone}
\langle F_{\alpha 0} \rangle = p\langle F_{\alpha 0} \rangle_1 ,
\end{equation}
then $\rho_2 $ is $\openone_S /N$ and $L(\rho_2 )$ is $\openone_S /N$, because $L(\openone_S )$ is $\openone_S $, so $\hat {L}(\langle F_{\alpha 0} \rangle_2 )$ is zero, and
\begin{equation}
\label{LFone}
\hat {L}(p\langle F_{\alpha 0} \rangle_1 ) = \hat {L}(\langle F_{\alpha 0} \rangle ) = p\hat {L}(\langle F_{\alpha 0} \rangle_1 )
\end{equation}
for all nonzero $\alpha $. When $\langle F_{\alpha 0} \rangle_2 $ is $-\langle F_{\alpha 0} \rangle_1 $ and $p$ is $1/2$, then $\langle F_{\alpha 0} \rangle $ is zero, so $\rho $ is $\openone_S /N$ and $L(\rho )$ is $\openone_S /N$, because $L(\openone_S )$ is $\openone_S $, so $\hat {L}(\langle F_{\alpha 0} \rangle )$ is zero and
\begin{equation}
\label{LFtwo}
\hat {L}(-\langle F_{\alpha 0} \rangle_1 ) = \hat {L}(\langle F_{\alpha 0} \rangle_2 ) = -\hat {L}(\langle F_{\alpha 0} \rangle_1 )
\end{equation}
for all nonzero $\alpha $. These equations are all we need to see that $\hat {L}$ extends to a linear map of the space of mean-value vectors.

\section{Inverse maps for fixed mean values}\label{five}

We consider the open quantum dynamics of $S$ described by a map $\Omega $ using Eqs.(\ref{eq:a3})-(\ref{eq:a4}) and (\ref{anyQ}) for fixed mean values. For each $L$, the different maps $\Omega $ for different $K$ all have inverses if and only if $L$ has an inverse. This is immediately evident from the observation that they all are one to one if and only if $L$ is. The inverse $\Omega^{-1} $ of $\Omega $ is that for any matrix $Q$ for $S$
\begin{equation}
  \label{eq112}
\Omega^{-1}(\Omega (Q)) =  L^{-1}(\Omega (Q) - K\Tr_S[\Omega (Q)])
\end{equation} 
because $\Tr_S [\Omega (Q)] = \Tr_S [Q]$. The inverse maps generally take density matrices to density matrices only in limited domains for limited sets of states of $S$. In this they are like the maps for the open quantum dynamics going forward \cite{jordan04,jordan05b}. They apply where they are meant to be used.

These statements are for all the maps $\Omega $ for fixed mean values. These maps are made to describe evolution of states of $S$ coming from dynamics described by a unitary operator $U$ for a larger system of $S$ and $R$ combined. Some of these maps $\Omega $ are completely positive; others are not, but $L$ always is completely positive. The conditions for an inverse are the same for maps that are not completely positive as for maps that are. Whether there is an inverse depends only on $L$, which depends only on $U$. It does not depend on the state of $R$ or on any mean values that involve $R$.

A necessary and sufficient condition for an inverse \cite{meLinearOperatorsTheorem72} is that $L$ does not map any nonzero matrix to zero. We can see from Eq.(\ref{Ldef}) that $L(Q^\dagger )$ is $L(Q)^\dagger $ for any matrix $Q$ for S. If $L(Q)$ is zero, so is $L(Q^\dagger )$. Hence the necessary and sufficient condition for an inverse is that $L$ does not map any nonzero Hermitian matrix to zero. We can see from Eq.(\ref{Ldef}) how this depends on $U$.

A necessary and sufficient condition for an inverse \cite{meLinearOperatorsTheorem72} is that $L$ maps the basis matrices $F_{\alpha  0}$ for $S$ to linearly independent matrices. In fact, the necessary and sufficient condition for an inverse is just that the matrices $L(F_{\alpha  0} )$ for positive $\alpha $ are linearly independent, because $L(\openone_S )$ is linearly independent of them, because $\Tr_S [L(F_{\alpha  0} )]$ is zero for positive $\alpha $ but $\Tr_S [L(\openone_S )]$ is not zero, because $L$ does not change the trace of $F_{\alpha  0}$, which we can see from Eq.(\ref{Ldef}). For each $L$, the matrices $L(F_{\alpha  0} )$ for positive $\alpha $ are the matrices $\Omega (F_{\alpha  0})$ of Eqs.(\ref{eq:l2}); they are the same for all the different maps $\Omega $ for different $K$.

Every matrix for $S$ is a linear combination of the $F_{\alpha  0}$. A necessary and sufficient condition for an inverse is that $L$ maps the linear space of all matrices for $S$ one to one onto itself. The alternative, the necessary and sufficient condition for no inverse, is that $L$ maps the $F_{\alpha  0}$ to linearly dependent matrices and maps the linear space of all matrices for $S$ into a subspace of itself that has fewer dimensions. Then there is a nonzero subspace of the linear space of all matrices for $S$ that is linearly independent of the subspace of all $L(Q)$ for matrices $Q$ for $S$. From Eq.(\ref{Ldef}), we can see that this gives a necessary and sufficient condition for an inverse in terms of $U$: that in the linear space of all matrices for $S$ and $R$ combined there is no nonzero subspace of matrices for $S$ that is linearly independent of the subspace of all $UQU^\dagger $ for matrices $Q$ for $S$.

In terms of $\hat {L}$, shown in Section IV to be a linear map of the space of mean-value vectors, a necessary and sufficient condition for an inverse \cite{meLinearOperatorsTheorem72} is that $\hat {L}$ does not map any nonzero mean-value vector to zero. For a qubit, this is the condition that the center is the only point in the Bloch sphere that $\hat {L}$ maps to the center. Since the components of a mean-value vector are the mean values $\langle F_{\alpha 0} \rangle $ for positive $\alpha $, the necessary and sufficient condition for an inverse is that the $\langle F_{\alpha 0} \rangle $ for positive $\alpha $ are not all zero after the dynamics if they were not all zero before. From Eq.(\ref{eq:f5}), we see that in terms of density matrices the necessary and sufficient condition for an inverse is that $L(\rho )$ is $\openone_S /N$ only if $\rho $ is $\openone_S /N$.

\section{Examples for fixed mean values} \label{six}

We consider two qubits, described by Pauli matrices $\psig 1$, $\psig 2$, $\psig 3$ for $S$ and $\pxi 1$, $\pxi 2$, $\pxi 3$ for $R$, so $F_{j0}$ is $\psig j$ and $F_{0k}$ is $\pxi k$, which implies $F_{jk}$ is $\psig j \pxi k$, for $j,k=1,2,3$. The density matrix for the two qubits is
\begin{equation}
  \label{eq:tq1}
  \Pi = \frac{1}{4} \left( \openone + \sum_{j=1}^3 \apsig{j} \psig j + \sum_{k=1}^3 \apxi{k} \pxi k + \sum_{j,k=1}^3 \apsigxi{j}{k} \psig j \pxi k \right)
\end{equation}
and the density matrix for $S$ is
\begin{equation}
  \label{eq:tq2}
  \rho = \Tr_R \Pi = \frac{1}{2} \left( \openone_S + \sum_{j=1}^3 \apsig{j} \psig j \right).
\end{equation}

Let
\begin{equation}
  \label{eq:int1}
  U = e^{-i \frac{1}{2} \gamma \psig 3 \pxi 3 }.
\end{equation}
We can easily compute
\begin{eqnarray}
  \label{eq:int2}
  U^{\dagger}\psig 1 U & = & \psig 1 e^{-i \gamma \psig 3 \pxi 3} \nonumber \\
 &=& \psig 1 \cos \gamma  - \psig 2 \pxi 3  \sin \gamma  
\end{eqnarray}
using the algebra of Pauli matrices, and similarly
\begin{eqnarray}
  \label{eq:int3}
  U^{\dagger}\psig 2 U &=& \psig 2 \cos \gamma  + \psig 1 \pxi 3 \sin \gamma  \nonumber \\
  U^{\dagger}\psig 3 U  &=& \psig 3 . 
\end{eqnarray}
Interchanging $U$ and $U^{\dagger}$ has the same effect as changing the sign of $\gamma $. Thus we see that 
\begin{eqnarray}
  \label{eq:int5}
  L(\openone_S ) & = & \openone_S  \nonumber \\
  L(\psig 1) & = & \psig 1 \cos \gamma \nonumber \\
   L(\psig 2) & = & \psig 2 \cos \gamma  \nonumber \\
    L(\psig 3) & = & \psig 3 .
\end{eqnarray}
The map $L$ is specified by these Eqs.(\ref{eq:int5}), or by the corresponding map of mean values
\begin{eqnarray}
  \label{Lmv}
  \hat {L}(\langle \psig 1 \rangle ) & = & \langle \psig 1 \rangle \cos \gamma  \nonumber \\
   \hat {L}(\langle \psig 2 \rangle ) & = & \langle \psig 2 \rangle \cos \gamma  \nonumber \\
    \hat {L}(\langle \psig 3 \rangle ) & = & \langle \psig 3 \rangle ,
\end{eqnarray}
or by
\begin{equation}
  \label{opsum}
   L(Q) =  \cos (\gamma /2)\, Q\, \cos (\gamma /2) + \psig 3 \sin (\gamma /2)\, Q\, \psig 3 \sin (\gamma /2)
\end{equation}
for any matrix $Q$ for $S$, which gives Eqs.(\ref{eq:int5}).

The necessary and sufficient condition for $L$ to have an inverse is that $\cos \gamma $ is not zero. Then the inverse map $L^{-1}$ is specified by
\begin{eqnarray}
  \label{Linv}
  L^{-1}(\openone_S ) & = & \openone_S  \nonumber \\
  L^{-1}(\psig 1) & = & \psig 1 /\cos \gamma \nonumber \\
   L^{-1}(\psig 2) & = & \psig 2 /\cos \gamma  \nonumber \\
    L^{-1}(\psig 3) & = & \psig 3 ,
\end{eqnarray}
\begin{eqnarray}
  \label{Lmvinv}
  \hat {L^{-1}}(\langle \psig 1 \rangle ) & = & \langle \psig 1 \rangle /\cos \gamma  \nonumber \\
   \hat {L^{-1}}(\langle \psig 2 \rangle ) & = & \langle \psig 2 \rangle /\cos \gamma  \nonumber \\
    \hat {L^{-1}}(\langle \psig 3 \rangle ) & = & \langle \psig 3 \rangle ,
\end{eqnarray}
or
\begin{equation}
  \label{invopsum}
   L^{-1}(Q) =  \pm \frac{\cos (\gamma /2)}{\sqrt{|\cos \gamma |}}\, Q\, \frac{\cos (\gamma /2)}{\sqrt{|\cos \gamma |}} \; \mp  \; \psig 3 \frac{\sin (\gamma /2)}{\sqrt{|\cos \gamma |}}\, Q\, \psig 3 \frac{\sin (\gamma /2)}{\sqrt{|\cos \gamma |}}
\end{equation}
with the upper signs for positive $\cos \gamma $ and the lower signs for negative $\cos \gamma $.
Clearly $L^{-1}$ is not completely positive. It does not map all positive matrices to positive matrices.

Here we considered the map $\Omega $ that is just $L$, with no $K$. From Eqs.(\ref{eq:int2}) and (\ref{eq:int3}), we can see that the map $\hat {\Omega }$ of mean values generally is
\begin{eqnarray}
  \label{mvmap}
\hat {\Omega }(\langle \psig 1 \rangle ) & = & \langle \psig 1 \rangle \cos \gamma  - \langle \psig 2 \pxi 3 \rangle \sin \gamma  \nonumber \\ 
\hat {\Omega }(\langle \psig 2 \rangle ) & = & \langle \psig 2 \rangle \cos \gamma  + \langle \psig 1 \pxi 3 \rangle \sin \gamma  \nonumber \\
\hat {\Omega }(\langle \psig 3 \rangle ) & = & \langle \psig 3 \rangle . 
\end{eqnarray}
It depends on $\langle \psig 2 \pxi 3 \rangle $ and $\langle \psig 1 \pxi 3 \rangle $ as well as on $U$. The map $\Omega $ is just $L$, with no $K$, when $\langle \psig 2 \pxi 3 \rangle $ and $\langle \psig 1 \pxi 3 \rangle $ are zero. In general, we need
\begin{eqnarray}
  \label{eq:int6}
  \Omega (\openone_S ) & = & \openone_S  + 2K \nonumber \\
   2K & = & [ - \apsigxi{2}{3} \Sigma_1
    + \apsigxi{1}{3} \Sigma_2 ]\sin \gamma 
\end{eqnarray}
to get the mean values in Eqs.(\ref{mvmap}) with
\begin{equation}
  \label{Ldm}
 L(\rho) + K = \Omega (\rho ) =  \frac{1}{2} [\Omega (\openone_S ) + \sum_{j=1}^3 \apsig{j} L(\psig j )].
\end{equation}

\section{Maps for fixed correlations}\label{seven}

Now we describe maps for fixed correlations \cite{stelmachovic01a}. We return to Eq.(\ref{Umv}), which gives the changes of the mean values $\langle F_{\alpha 0} \rangle $ for one state of $S$. From there we take a different route to describe the open quantum dynamics of $S$ with a single map that applies to a set of states of $S$. The result is a different map, sometimes an infinite number of different maps, replacing the map $\Omega $ that we obtained for fixed mean values.

We use the correlations
\begin{equation}
\label{corr}
\Gamma _{\mu \nu } = \langle F_{\mu \nu } \rangle  - \langle F_{\mu  0} \rangle \langle F_{0 \nu } \rangle 
\end{equation}
for positive $\mu$, $\nu$. We define a map $\hat {\Phi }$ of the mean values $\langle F_{\alpha 0} \rangle $ for positive $\alpha $ by using Eq.(\ref{corr}) to substitute for $\langle F_{\mu \nu } \rangle $ in Eq.(\ref{Umv}) and letting
\begin{eqnarray}
  \label{Umvc}
 \hat {\Phi }(\langle F_{\alpha 0} \rangle ) & = & \langle U^{\dagger}F_{\alpha 0} U\rangle  \nonumber \\
& = & \sum_{\mu  =1}^{N^2-1}t_{\alpha 0 \,;\, \mu 0}\langle F_{\mu 0}\rangle  + \sum_{\nu =1}^{M^2-1}t_{\alpha 0 \,;\, 0 \nu }\langle F_{0 \nu }\rangle \nonumber \\
& & + \sum_{\mu =1}^{N^2-1} \sum_{\nu =1}^{M^2-1}t_{\alpha 0 \,;\, \mu  \nu }\langle F_{\mu  0} \rangle \langle F_{0 \nu } \rangle  + \sum_{\mu =1}^{N^2-1} \sum_{\nu =1}^{M^2-1}t_{\alpha 0 \,;\, \mu  \nu }\Gamma_{\mu \nu }
\end{eqnarray}
for fixed mean values $\langle F_{0 \nu }\rangle $ or density matrix
\begin{equation}
  \label{rhoR}
  \rho_R = \frac{1}{M}[\openone_R + \sum_{\nu =1}^{M^2-1} \langle F_{0 \nu } \rangle F_{0 \nu }]
\end{equation}
for the state of $R$, and fixed correlations $\Gamma _{\mu \nu }$.
We consider the $\langle F_{0 \nu }\rangle $ and $\Gamma _{\mu \nu }$ that participate in Eq.(\ref{Umvc}) to be map parameters. They describe the effect of the dynamics of the larger system of $S$ and $R$ combined that drives the evolution of $S$. There may be $\langle F_{0 \nu }\rangle $ and $\Gamma _{\mu \nu }$ that are not map parameters; they have no effect on the change of states of $S$ because in Eq.(\ref{Umvc}) they are with $t_{\alpha 0 \,;\, 0 \nu }$ or $t_{\alpha 0 \,;\, \mu  \nu }$ that are zero. Which $\langle F_{0 \nu }\rangle $ and $\Gamma _{\mu \nu }$ are map parameters depends on $U$.

This map $\hat {\Phi }$ applies to all the states of $S$ described by mean values $\langle F_{\alpha 0} \rangle $ that are compatible with the fixed map parameters $\langle F_{0 \nu } \rangle $ and $\Gamma _{\mu \nu }$ in describing a possible initial state of $S$ and $R$ combined. We call this set of states of $S$ the compatibility domain of the map $\Phi $. It may be smaller than the compatibility domain of a corresponding map $\Omega $ for fixed mean values, because for $\Phi $ the $\langle F_{\alpha 0} \rangle $ have to be compatible with a larger number of fixed parameters. Generally there is an added parameter for $\Phi $ when there is a $\mu $ and a $\nu $ for which $t_{\alpha 0 \,;\, 0 \nu }$ is zero for all $\alpha $ but $t_{\alpha 0 \,;\, \mu  \nu }$ is not. Then $\langle F_{\mu \nu } \rangle $ is a parameter for $\Omega $ but $\langle F_{0 \nu } \rangle $ is not, and both $\Gamma _{\mu \nu }$ and $\langle F_{0 \nu } \rangle $ are parameters for $\Phi $ in all but the exceptional cases where there is a cancellation of $t_{\alpha 0 \,;\, \mu  \nu }\langle F_{0 \nu } \rangle $ with $t_{\alpha 0 \,;\, \mu 0}$. In Sections VI and IX we will see examples where $\Omega $ depends on two parameters and $\Phi $ depends on three.

For each state of $S$ in the compatibility domain, the state of $S$ and $R$ is described by a density matrix 
\begin{eqnarray}
  \label{PiPhi}
  \Pi_\Phi  & = & \frac{1}{NM}[ \openone + \sum_{\alpha =1}^{N^2-1} \langle F_{\alpha 0} \rangle F_{\alpha 0} + \sum_{\nu  =1}^{M^2-1} \langle F_{0 \nu } \rangle F_{0 \nu } \nonumber \\
& & + \sum_{\mu =1}^{N^2-1}\sum_{\nu  =1}^{M^2-1} \langle F_{\mu 0} \rangle \langle F_{0 \nu } \rangle F_{\mu \nu }
+ \sum_{\mu =1}^{N^2-1}\sum_{\nu  =1}^{M^2-1} \Gamma_{\mu \nu } F_{\mu \nu }]
\end{eqnarray}
obtained by using Eq.(\ref{corr}) to substitute for $\langle F_{\mu \nu } \rangle $ in Eq.(\ref{eq:f4}).

The mean value $\langle A \rangle = \Tr_S[A\rho ]$ for a matrix $A$ for $S$ is changed to
\begin{equation}
  \label{eq:a2}
 \langle U^{\dagger}A U \rangle = \Tr[U^{\dagger}A U\Pi_\Phi ] = \Tr_S[ A \; \Tr_R[U \Pi_\Phi  U^{\dagger}]]
\end{equation}
so now the Schr\"{o}dinger picture for $S$ is that the density matrix $\rho$ for $S$ is changed to\footnotemark[1]
\begin{equation}
  \label{PhiDC}
  \Phi (\rho ) = \Tr_R[ U \Pi_\Phi U^{\dagger}] = D(\rho) + C
\end{equation}
where
\begin{equation}
  \label{Ddef}
  D(Q) = \Tr_R[ U \, Q \,\rho _R \, U^{\dagger}]  
\end{equation}
\begin{equation}
\label{Cdef}
  C = \Tr_R[ U(\Pi_\Phi  - \rho  \, \rho _R )U^{\dagger}] = \frac{1}{NM}  \sum_{\mu =1}^{N^2-1}\sum_{\nu  =1}^{M^2-1} \Gamma _{\mu \nu } \Tr_R[UF_{\mu \nu }U^{\dagger}] .
\end{equation}

The $D$ part is a completely positive linear map that applies to any matrix $Q$ for $S$, density matrix or not. The map $D$ depends on $U$ and on the state of $R$ represented by the density matrix $\rho_R$. It does not depend on the correlations $\Gamma _{\mu \nu }$ between $S$ and $R$. The matrix $C$ is the only part of $\Phi (\rho )$ that depends on the correlations $\Gamma _{\mu \nu }$. It also depends on $U$, but it does not depend on the state of $S$ described by the mean values $\langle F_{\alpha 0} \rangle$ or density matrix $\rho $ or on the state of $R$ described by the mean values $\langle F_{0 \nu } \rangle$ or density matrix $\rho_R $. 

Now the map parameters are the $\langle F_{0 \nu } \rangle$ and $\Gamma _{\mu \nu }$. Different $\langle F_{0 \nu } \rangle$ and $\Gamma _{\mu \nu }$ specify different maps. Each map $\Phi  $ applies to different states of $S$ described by different $\langle F_{\alpha 0} \rangle $. For each map $\Phi  $ there is one density matrix $\rho_R $ and one $N$$\times$$N$ matrix $C$.

The map $\Phi  $ of density matrices $\rho $ described by Eqs.(\ref{PhiDC})-(\ref{Cdef}) extends to a linear map of all matrices for $S$. The map of the basis matrices $F_{\alpha 0}$ to
\begin{equation}
  \label{Phibasis}
  \Phi  (\openone_S ) = \openone_S + NC, \; \; \; \; \quad \Phi (F_{\alpha 0}) = D(F_{\alpha 0})
\end{equation}
for positive $\alpha $ is equivalent to Eqs.(\ref{PhiDC})-(\ref{Cdef}) for a set of $\rho $ described by Eq.(\ref{eq:f5}) with variable $\langle F_{\alpha 0} \rangle$. Since $C$ is the same for all $\rho$, it cannot come from the terms with variable coefficients $\langle F_{\alpha 0} \rangle$; it can only be part of $\Phi (\openone_S )$. For each matrix $Q$ for $S$, density matrix or not, the map is that
\begin{equation}
  \label{PhiQ}
  \Phi (Q) = D(Q) + C\Tr_S (Q).
\end{equation}

Let $E$ be the map defined by
\begin{equation}
  \label{Edef}
 E(Q) = D(Q) - D(\openone_S )\Tr_S (Q)/N + \openone_S \Tr_S (Q)/N
\end{equation}
for all matrices $Q$ for $S$. Then $E$ is a linear map, and $E(\openone_S )$ is $\openone_S $, so the corresponding map $\hat {E}$ of mean values is a linear map of the space of mean-value vectors, as was shown for $\hat {L}$ in Section IV.

The maps $\Phi  $ include the familiar completely positive maps that have long been used for initial states with no correlations. When the correlations $\Gamma _{\mu \nu }$ are zero, $C$ is zero and Eqs.(\ref{PhiDC}) and (\ref{Ddef}) are the familiar equations that define the familiar completely positive maps \cite{nielsen00}. When maps defined with fixed mean values are used, the familiar completely positive maps are separate different maps, alternative maps that are options when there are no correlations in the map parameters \cite{jordan04}.

 \section{Inverse maps for fixed correlations}\label{eight}

Inverses of maps for fixed correlations parallel those for fixed mean values. For each $D$, the different maps $\Phi  $ for different $C$ all have inverses if and only if $D$ has an inverse. The inverses $\Phi ^{-1} $ are described by Eq.(\ref{eq112}) with $\Omega $, $L$ and $K$ replaced by $\Phi $, $D$ and $C$. Again, the inverse maps generally take density matrices to density matrices only in limited domains for limited sets of states of $S$. In this they are like the maps for the open quantum dynamics going forward. They apply where they are meant to be used.

These statements are for all the maps $\Phi $ for fixed correlations. These maps are made to describe evolution of states of $S$ coming from dynamics described by a unitary operator $U$ for a larger system of $S$ and $R$ combined. Some of these maps $\Phi $ are completely positive; others are not, but $D$ always is completely positive. The conditions for an inverse are the same for maps that are not completely positive as for maps that are. Whether there is an inverse depends only on $D$, which depends only on $U$ and on the state of $R$ described by the density matrix $\rho_R $. It does not depend on the correlations, or absence of correlations, between $S$ and $R$.

A necessary and sufficient condition for an inverse is that $D$ does not map any nonzero Hermitian matrix to zero, or that the matrices $D(F_{\alpha  0} )$ for positive $\alpha $ are linearly independent. For each $D$, the $D(F_{\alpha  0} )$ for positive $\alpha $ are the $\Phi (F_{\alpha  0})$ of Eqs.(\ref{Phibasis}); they are the same for all the different maps $\Phi $ for different $C$. The discussion for fixed mean values in Section V applies almost unchanged with $L$ replaced by $D$. One change is that $D(\openone_S )$ is generally not $\openone_S $. In terms of the map of mean-value vectors, a necessary and sufficient condition for an inverse is that $\hat {E}$ does not map any nonzero mean-value vector to zero, which means that $\hat {D}$ does not map the zero mean-value vector and another mean-value vector both to the same vector, which means that $D$ does not map $\openone_S /N$ and another density matrix both to the same density matrix.\footnotemark[1] For a qubit, this is the condition that $\hat {D}$ does not map the center and another point in the Bloch sphere both to the same point. For maps for fixed correlations, there is no necessary and sufficient condition for an inverse in terms of $U$ alone, as there is for maps with fixed mean values; we will see an example where the condition for an inverse involves both $U$ and the state of $R$.

\section{Examples for fixed correlations}\label{nine}

Again, we consider two qubits, described by Pauli matrices $\psig 1$, $\psig 2$, $\psig 3$ and $\pxi 1$, $\pxi 2$, $\pxi 3$, and an open quantum dynamics of the $\Sigma $ qubit that comes from the dynamics described by the unitary operator $U$ of Eq.(\ref{eq:int1}) for the two qubits. Now we let each map be specified by fixed mean values $\langle \Xi_k \rangle $ and fixed correlations
\begin{equation}
\label{corrjk} 
\Gamma_{jk} = \langle \Sigma_j \Xi_k \rangle  - \langle \Sigma_j \rangle \langle \Xi_k \rangle 
\end{equation}
for j, k = 1, 2, 3. Taking mean values in Eqs.(\ref{eq:int2}) and (\ref{eq:int3}) and using Eqs.(\ref{corrjk}) to substitute for $\langle \psig 2 \pxi 3 \rangle $ and $\langle \psig 1 \pxi 3 \rangle $, gives the map of mean values
\begin{eqnarray}
  \label{Dmvinv}
\hat {\Phi }(\langle \psig 1 \rangle ) & = & \langle \psig 1 \rangle \cos \gamma  - \langle \psig 2 \rangle \langle \pxi 3 \rangle \sin \gamma  - \Gamma_{23} \sin \gamma \nonumber \\ 
\hat {\Phi }(\langle \psig 2 \rangle ) & = & \langle \psig 2 \rangle \cos \gamma  + \langle \psig 1 \rangle \langle \pxi 3 \rangle \sin \gamma  + \Gamma_{13} \sin \gamma \nonumber \\
\hat {\Phi }(\langle \psig 3 \rangle ) & = & \langle \psig 3 \rangle . 
\end{eqnarray}
From calculations like those in Eqs.(\ref{eq:int2}) and (\ref{eq:int3}), with the sign of $\gamma $ changed to interchange $U$ and $U^\dagger $, we find that
\begin{eqnarray}
  \label{Dsig}
  D(\openone_S ) & = & \openone_S  \nonumber \\
  D(\psig 1) & = & \psig 1 \cos \gamma  + \psig 2 \langle \pxi 3 \rangle \sin \gamma \nonumber \\
   D(\psig 2) & = & \psig 2 \cos \gamma  - \psig 1 \langle \pxi 3 \rangle \sin \gamma \nonumber \\
    D(\psig 3) & = & \psig 3 \\
C & = & \frac{1}{2}(\Gamma_{13}\psig 2  - \Gamma_{23}\psig 1 )\sin \gamma .
\end{eqnarray}
We can see that these give a map $\Phi $ that gives the same changes of density matrices $\rho $ as the map $\hat {\Phi }$ of mean values described by Eqs.(\ref{Dmvinv}). The map $\Phi $ is completely positive only if $\Gamma_{13}$ and $\Gamma_{23}$ are both zero; only then is the matrix 
\begin{equation}
  \label{Phipos}
  \Phi (\frac{1}{2}[\openone_S + \psig 3 ]) = \frac{1}{2}[\openone_S + \psig 3 + \Gamma_{13}\psig 2 \sin \gamma - \Gamma_{23} \psig 1 \sin \gamma ]
\end{equation}
a positive matrix.

The condition for an inverse is that the determinant
\begin{equation}
  \label{det}
\det = (\cos \gamma )^2 + \langle \pxi 3 \rangle^2(\sin \gamma )^2
\end{equation}
is not zero, which means that $\cos \gamma $ and $\langle \pxi 3 \rangle $ are not both zero. If $\cos \gamma $ is zero, the condition for an inverse is that $\langle \pxi 3 \rangle $ is not zero. Whether there is an inverse depends on both $U$ and the state of $R$. The inverse map of basis matrices is
\begin{eqnarray}
  \label{Dsiginv}
  D^{-1}(\openone_S ) & = & \openone_S  \nonumber \\
  D^{-1}(\psig 1) & = & \psig 1 \frac{\cos \gamma }{\det} - \psig 2 \frac{\langle \pxi 3 \rangle \sin \gamma }{\det} \nonumber \\
   D^{-1}(\psig 2) & = & \psig 2 \frac{\cos \gamma }{\det} + \psig 1 \frac{\langle \pxi 3 \rangle \sin \gamma }{\det} \nonumber \\
    D^{-1}(\psig 3) & = & \psig 3 .
\end{eqnarray}
The corresponding map of mean values is
\begin{eqnarray}
  \label{Dmvinv2}
\hat {D^{-1}}(\langle \psig 1 \rangle ) & = & \langle \psig 1 \rangle \frac{\cos \gamma }{\det} + \langle \psig 2 \rangle \frac{\langle \pxi 3 \rangle \sin \gamma }{\det}  \nonumber \\ 
\hat {D^{-1}}(\langle \psig 2 \rangle ) & = & \langle \psig 2 \rangle \frac{\cos \gamma }{\det} - \langle \psig 1 \rangle \frac{\langle \pxi 3 \rangle \sin \gamma }{\det}  \nonumber \\
\hat {D^{-1}}(\langle \psig 3 \rangle ) & = & \langle \psig 3 \rangle . 
\end{eqnarray}
This gives
\begin{equation}
  \label{Dinvsq}
[\hat {D^{-1}}(\langle \psig 1 \rangle )]^2  +  [\hat {D^{-1}}(\langle \psig 2 \rangle )]^2 = \frac{\langle \psig 1 \rangle^2  + \langle \psig 2 \rangle^2}{\det}
\end{equation}
which shows that $D^{-1}$ is not completely positive unless $\langle \pxi 3 \rangle $ is $1$ or $\sin\gamma $ is zero. From Eqs.(\ref{Dmvinv}) we see that if $\sin \gamma $ is zero, then $\hat {\Phi }$ can either do nothing or just change the signs of $\langle \psig 1 \rangle $ and $\langle \psig 2 \rangle $. If $\langle \pxi 3 \rangle $ is $1$, then the state of $R$ is a pure state, so $\Pi_\Phi $ is a product $\rho \rho_R $ for any state of $S$ in the compatibility domain, which means there are no correlations, and the map described by $\Phi $ or $\hat {\Phi }$ is just rotation by $\gamma $ around the $z$ axis.

\section{Dynamics disconnection}\label{ten}

For the larger system of $S$ and $R$ combined, we can consider both the dynamics described by $U$ that takes states forward for an interval of time, and the reverse dynamics described by $U^\dagger $ that takes states backwards through the same interval. We consider maps for constant mean values first. In our example for constant mean values, we assume now that $\langle \psig 2 \pxi 3 \rangle $ and $\langle \psig 1 \pxi 3 \rangle $ are zero at the start of the interval, so the map $\Omega $ that describes the evolution of states of $S$ going forward through the interval is just $L$, with no $K$. Since the change from $U$ to $U^\dagger $ is just the change of sign of $\gamma $, the maps that describe the changes of states of $S$ going backwards through the interval have the same $L$, described by Eqs.(\ref{eq:int5}) or (\ref{opsum}), and have $K$ matrices described by Eqs.(\ref{eq:int6}), which give maps of mean values described by Eqs.(\ref{mvmap}), with $\sin \gamma $ changed to $-\sin \gamma $ and $\langle \psig 2 \pxi 3 \rangle $ and $\langle \psig 1 \pxi 3 \rangle $ changed to the values $\langle \psig 2 \pxi 3 \rangle '$ and $\langle \psig 1 \pxi 3 \rangle '$ at the end of the interval, which, from calculations like those of Eqs.(\ref{eq:int2}) and (\ref{eq:int3}), can be seen to be
\begin{eqnarray}
  \label{mvmap2}
\langle \psig 2 \pxi 3 \rangle' & = & \langle \psig 1 \rangle  \sin \gamma  \nonumber \\ 
\langle \psig 1 \pxi 3 \rangle ' & = & -\langle \psig 2 \rangle \sin \gamma  
\end{eqnarray}
with $\langle \psig 1 \rangle $ and $\langle \psig 2 \rangle $ mean values at the start of the interval. After the changes forward and backward, the mean values $\langle \psig j \rangle $ are returned, as we know they must be, to
\begin{eqnarray}
  \label{mvreturn}
\langle \psig 1 \rangle '' & = & \hat {L}(\langle \psig 1 \rangle )\cos \gamma  + \langle \psig 2 \pxi 3 \rangle '\sin \gamma  \nonumber \\ 
 & = & \langle \psig 1 \rangle (\cos \gamma )^2 + \langle \psig 1 \rangle (\sin \gamma )^2 = \langle \psig 1 \rangle  \nonumber \\ 
\langle \psig 2 \rangle '' & = & \hat {L}(\langle \psig 2 \rangle )\cos \gamma  - \langle \psig 1 \pxi 3 \rangle '\sin \gamma  \nonumber \\
 & = & \langle \psig 2 \rangle (\cos \gamma )^2 + \langle \psig 2 \rangle (\sin \gamma )^2 = \langle \psig 2 \rangle  \nonumber \\
\langle \psig 3 \rangle '' & = & \hat {L}(\langle \psig 3 \rangle ) = \langle \psig 3 \rangle. 
\end{eqnarray}
The maps that describe the changes of states of $S$ going backwards in time are different for different values of $\langle \psig 1 \rangle $ and $\langle \psig 2 \rangle $; their $K$ matrices are different because the $\langle \psig 2 \pxi 3 \rangle '$ and $\langle \psig 1 \pxi 3 \rangle '$ given by Eqs.(\ref{mvmap2}) are different. These maps generally are different for states of $S$ that are different at the start of the interval. They do not comprise a single inverse map for all the states of $S$ that are mapped forward.

The inverse map $L^{-1}$ in our example is not obtained from $U^\dagger $ and equations like Eqs.(\ref{eq:a3})-(\ref{eq:a4}). It is not tied to the dynamics the way $L$ is. In fact, the map $L^{-1}$ can not be obtained the way $L$ is, as a map for fixed mean values, from any dynamics described by any unitary operator for any states of a larger system of $S$ and $R$ combined. There are two ways to see this. One is that if $L^{-1}$ were obtained that way, it would have to be completely positive, because its $K$ would be zero, because $L^{-1}(\openone_S )$ is $\openone_S $; but $L^{-1}$ is not completely positive. Another way to see that $L^{-1}$ is not obtained from dynamics, described next, will apply to our example for fixed correlations as well.

\section{No dynamics possible}\label{eleven}

There is no way that an inverse can be obtained from dynamics for any unital trace-preserving completely-positive map that describes truly open dynamics. A unital map is a map that takes the unit matrix to the unit matrix, so it maps the density matrix for the completely mixed state to itself.
\vspace{0.6cm}

\noindent\textit{Theorem.} If a completely positive map preserves the trace, is unital, and is not simply a unitary transformation for the subsystem $S$ alone, then it does not have an inverse that is obtained from any dynamics described by any unitary operator for any larger system of $S$ combined with another system $R$.
\vspace{0.6cm}

\noindent\textit{Proof.} A trace-preserving completely positive map can be obtained from dynamics as a map for fixed correlations that are zero \cite{nielsen00}. It is a map $D$ described by Eq.(\ref{Ddef}) for some $R$ and $U$. If the map is unital, then $D(\openone_S )$ is $\openone_S $, and the map $E$ defined by Eq.(\ref{Edef}) is $D$, so $\hat {D}$ is a linear map of the space of mean-value vectors.\footnotemark[1] If the map has an inverse $D^{-1}$, then $\hat {D^{-1}}$ is the inverse of $\hat {D}$ and is a linear map of the space of mean-value vectors. The density matrix $\rho $ for a state of $S$, described by Eq.(\ref{eq:f5}), is changed by $D$ to
\begin{equation}
  \label{rhot}
  D(\rho ) = \frac{1}{N}[ \openone_S + \sum_{\alpha=1}^{N^2-1} \hat {D}(\langle F_{\alpha 0} \rangle )F_{\alpha 0}]
\end{equation}
with $\hat {D}(\langle F_{\alpha 0} \rangle ) = \langle U^{\dagger}F_{\alpha 0} U\rangle $ calculated from Eq.(\ref{Umv}). Since $\hat {D}$ is a linear map of the space of mean-value vectors, each $\hat {D}(\langle F_{\alpha 0} \rangle )$ given by Eq.(\ref{Umv}) is a linear combination of the $\langle F_{\mu 0} \rangle$ alone. This implies that
\begin{equation}
  \label{eq:p1}
  \Tr \left[ \left( D(\rho )\right)^2 \right] \leq \Tr \left[ \rho^2 \right].
\end{equation}
To see this, let
\begin{equation}
  \label{eq:p5}
  \chi_{\alpha \beta} = \langle F_{\alpha 0} \rangle \quad {\mbox{for}} \quad \alpha =1,2,\ldots N^2-1\,, \quad \beta=0
\end{equation}
and $\chi_{\alpha \beta}=0$ for other $\alpha$, $\beta$, and let
\begin{equation}
  \label{eq:p6}
  \hat {D}(\chi_{\mu \nu}) = \sum_{\alpha=0}^{N^2-1} \sum_{\beta=0}^{M^2-1} t_{\mu \nu \,;\, \alpha \beta} \chi_{\alpha \beta}.
\end{equation}
Then $\hat {D}(\langle F_{\mu 0} \rangle )$ is $\hat {D}(\chi_{\mu 0})$ for positive $\mu $, because 
$t_{\mu 0\,;\, 00}$ is zero, and
\begin{equation}
  \label{eq:p7}
  \sum_{\mu=1}^{N^2-1} \left( \hat {D}(\langle F_{\mu 0}\rangle ) \right)^2 \leq \sum_{\mu=0}^{N^2-1} \sum_{\nu=0}^{M^2-1}\left( \hat {D}(\chi_{\mu \nu}) \right)^2 = \sum_{\alpha=0}^{N^2-1} \sum_{\beta=0}^{M^2-1} \left(\chi_{\alpha \beta} \right)^2 = \sum_{\alpha=1}^{N^2-1} \langle F_{\alpha 0} \rangle^2
\end{equation}
because the $t_{\mu \nu \,;\, \alpha \beta}$ are the elements of a real orthogonal matrix. This proves the inequality (\ref{eq:p1}). If $D^{-1}$ comes from dynamics (with both $R$ and $U$ generally different than for $D$), we can see similarly that
\begin{equation}
  \label{eq:p1b}
  \Tr \left[ \rho^2 \right] \leq \Tr \left[ \left( D(\rho )\right)^2 \right].
\end{equation}

Pure states are mapped to pure states. For each vector $|\psi \rangle $ of length one, there is a vector $|\psi^D \rangle $ of length one such that
\begin{equation}
  \label{vecmap}
  D(|\psi \rangle \langle \psi |) = |\psi^D \rangle \langle \psi^D |.
\end{equation}
Absolute values of inner products of vectors are not changed: if
\begin{equation}
  \label{rho22}
  \rho  = \frac{1}{2}|\psi \rangle \langle \psi | + \frac{1}{2}|\phi  \rangle \langle \phi  |
\end{equation}
for two vectors $|\psi \rangle $ and $|\phi  \rangle $ of length one, then
\begin{equation}
  \label{avip}
  \frac{1}{2} + \frac{1}{2}|\langle \psi |\phi \rangle |^2 = \Tr \left[ \rho^2 \right] = \Tr \left[ \left( D(\rho )\right)^2 \right] = \frac{1}{2} + \frac{1}{2}|\langle \psi^D |\phi^D \rangle |^2.
\end{equation}
The trace of a product of Hermitian matrices is not changed: for one-dimensional projection operators $|\psi \rangle \langle \psi |$ and $|\phi  \rangle \langle \phi  |$ for vectors $|\psi \rangle $ and $|\phi  \rangle $ of length one,
\begin{equation}
  \label{trip}
  \Tr \left[ |\psi \rangle \langle \psi ||\phi  \rangle \langle \phi  | \right] = |\langle \psi |\phi \rangle |^2 = |\langle \psi^D |\phi^D \rangle |^2 = \Tr \left[ D(|\psi \rangle \langle \psi |)D(|\phi  \rangle \langle \phi  |) \right]
\end{equation}
and for any Hermitian operators $F$ and $G$ there are spectral decompositions
\begin{equation}
  \label{specdecom}
  F = \sum_j f_j |\psi_j \rangle \langle \psi_j |, \quad G = \sum_k g_k |\phi _k \rangle \langle \phi _k |
\end{equation}
with real numbers $f_j $ and $g_k $ and vectors $|\psi_j \rangle $ and $|\phi _k \rangle $ of length one, which give
\begin{eqnarray}
  \label{trprod}
\Tr \left[ FG\right] & = & \sum_{jk} f_j \, g_k\Tr \left[ |\psi_j \rangle \langle \psi_j ||\phi_k  \rangle \langle \phi_k  | \right] \nonumber \\
& = & \sum_{jk} f_j \, g_k\Tr \left[ D(|\psi_j \rangle \langle \psi_j |)D(|\phi_k  \rangle \langle \phi_k  |) \right] = \Tr \left[ D(F)D(G)\right].
\end{eqnarray}
Equation (\ref{eq:f1}) holds for the $D(F_{\mu 0} )$ as well as for the $F_{\mu 0} $. This implies that the $N^2 $ matrices $D(F_{\mu 0} )$ are linearly independent, so every matrix for $S$ is a linear combination of the $D(F_{\mu 0} )$. The linear space of all matrices for $S$ is mapped one-to-one onto itself. The set of all pure states is mapped one-to-one onto itself. Then Wigner's theorem says that the map is made by either a linear unitary operator or an antilinear antiunitary operator on the space of state vectors \cite{WignerGroupTheory,Bargmann64}. An antilinear antiunitary operator is not possible \cite{kraus83}. The map is a unitary transformation for $S$ alone. This completes the proof of the theorem.

In particular, in our examples, neither $L^{-1}$ nor $D^{-1}$ generally can be obtained from any dynamics described by any unitary operator for any states of a larger system of $S$ combined with another system $R$. The only exceptions are when $L$ is the identity map and when $D$ is rotation of the $\Sigma $ qubit around the $z$ axis.

\section{Discussion}\label{twelve}

Either way the map is defined, with only fixed mean values as map parameters or with fixed correlations, the map gives a correct description of the change in time of every state of $S$ in its compatibility domain; the map describes the change that comes from the unitary dynamics in the situation described by the map parameters in the larger system of $S$ and $R$ combined. Our examples show that as a whole the map defined one way can be quite different from the map defined the other way. The map $D$ described by Eqs.(\ref{Dsig}) contains a rotation that is not in the map $L$ described by Eqs.(\ref{eq:int5}) or (\ref{Lmv}). The compatibility domains of the two kinds of maps are different, and for states that are in both of their compatibility domains, maps $\Omega $ and $\Phi $ generally are different. In our examples, maps $\Omega $ and $\Phi $ give the same result only for the one set of values of $\langle \psig 1 \rangle $ and $\langle \psig 2 \rangle $ that make $\hat {\Omega }(\langle \psig 1 \rangle )$ and $\hat {\Omega }(\langle \psig 2 \rangle )$ described by Eqs.(\ref{mvmap}) the same as $\hat {\Phi }(\langle \psig 1 \rangle )$ and $\hat {\Phi }(\langle \psig 2 \rangle )$ described by Eqs.(\ref{Dmvinv}); these are the $\langle \psig 1 \rangle $ and $\langle \psig 2 \rangle $ that are solutions of  Eqs.(\ref{corrjk}) for the fixed parameters $\langle \psig 2 \pxi 3 \rangle $ and $\langle \psig 1 \pxi 3 \rangle $ of $\Omega $ and the fixed parameters $\langle \pxi 3 \rangle $, $\Gamma_{23} $ and $\Gamma_{13} $ of $\Phi $.

Which kind of map is the better tool to use may depend on what the job to be done is. In particular, one compatibility domain may extend better than the other to include states of interest. The differences between the two kinds of maps do not reflect physical properties of the dynamics. Far from it; they are differences between two ways of describing the effect of the same dynamics of $S$ and $R$ on a set of states of $S$, either a set of states compatible with fixed mean values or a set of states compatible with fixed correlations.

Whether a map has an inverse is a property of the map that may be closely tied to the dynamics. The conditions for an inverse are very similar for the two kinds of maps. For maps defined with only fixed mean values as map parameters, we saw in Section V that a necessary and sufficient condition for an inverse can be stated just in terms of the unitary operator that describes the dynamics of $S$ and $R$. For maps defined with fixed correlations, our examples in Section IX show that the dynamics alone does not always determine whether a map has an inverse; it can depend on the state of $R$ as well. The conditions for an inverse are the same for maps that are not completely positive as for maps that are.

The dynamics is generally less closely tied to what the inverse is. We saw in Section X that when a map comes from dynamics, the inverse map generally does not come from the reversed dynamics. It might not come from any dynamics at all. We showed in Section XI that if a trace-preserving completely positive map is unital, it can not have an inverse that is obtained from any dynamics described by any unitary operator for any states of a larger system. When a map is an operation that can be done to the system, its inverse might not be.

Many maps that are not completely positive describe operations that can be done to a system through unitary dynamical interactions with other systems. Many maps do not. Many inverses of maps that do, do not. All trace-preserving completely positive maps do \cite{nielsen00}.
 
We can appreciate how quantum error correction works with completely positive maps. Using only completely positive maps avoids maps that can not be realized physically. The recovery map can be completely positive because it is not a full inverse. It is an inverse only on a subspace of the code space \cite{nielsen00}. If a trace-preserving completely positive map has a full inverse that is completely positive, the map comes from a unitary operator on the system alone, without any interaction with another system \cite{Nielsen98}. Completely positive maps do not describe everything that can happen in open quantum systems, but they are the maps that are most simply, completely, and certainly tied to dynamics.

\section*{ACKNOWLEDGMENT}

I am grateful to Anil Shaji and George Sudarshan for very extensive discussions and help with this subject over the past several years.

\bibliography{ncp}
\end{document}